\documentclass[12pt]{article}
\usepackage[latin1]{inputenc}
\usepackage{enumerate}
\usepackage{natbib}
\usepackage{amsmath}
\usepackage{amssymb}
\usepackage{indentfirst}
\usepackage[T1]{fontenc}
\usepackage{graphicx}
\usepackage{graphics}
\usepackage{epstopdf} 
\usepackage{hyperref} %produz referência clicáveis no documento.
\usepackage{ae}
\usepackage{url}
\usepackage{caption2}
\newtheorem{theo}{Theorem}

\usepackage{vmargin}
\setmarginsrb{2.5cm}{2.cm}{2.5cm}{2.cm}{1.cm}{1mm}{1cm}{1cm}
\title{Doubly Robust-Based Generalized Estimating Equations for the Analysis of Longitudinal Ordinal Missing Data}
\author{José Luiz P. da Silva, Enrico A. Colosimo, Fábio N. Demarqui\\Departament of Statistics\\Universidade Federal de Minas Gerais}
\begin{document}
\maketitle
\begin{abstract}
Generalized Estimation Equations (GEE) are a well-known method for the analysis of non-Gaussian longitudinal data. This method has computational simplicity and marginal parameter interpretation. However, in the presence of missing data, it is only valid under the strong assumption of missing completely at random (MCAR). Some corrections can be done when the missing data mechanism is missing at random (MAR): inverse probability weighting (WGEE) and multiple imputation (MIGEE). In order to obtain consistent estimates, it is necessary the correct specification of the weight model for WGEE or the imputation model for the MIGEE. A recent method combining ideas of these two approaches has doubly robust property. For consistency, it requires only the weight or the imputation model to be correct. In this work it is assumed a proportional odds model and it is proposed a doubly robust estimator for the analysis of ordinal longitudinal data with intermittently missing response and covariate under the MAR mechanism. Simulation results revealed better performance of the proposed method compared to WGEE and MIGEE. The method is applied to a data set related to Analgesia Pain in Childbirth study.
\end{abstract}

\section{Introduction}\label{um_intro}

The Generalized Estimating Equation (GEE) method \citep{liang1986} is one of the most popular approaches for the analysis of non-Gaussian correlated data. Its main advantage resides in the fact that one is only required to specify correctly the mean structure of the response for the parameter estimator to be consistent and asymptotically normal. In its basic formulation the association parameters among repeated measures were taken as nuisance parameters. GEE method has computational simplicity (there is no need of dealing with complex, and in some cases, intractable likelihoods) and it further allows populational-averaged interpretation of the parameter of interest. 

It is very common for sets of longitudinal data to be incomplete, in the sense that not all planned observations are actually observed. This problem is pervasive in longitudinal data because nonresponse can occur any time from the beginning of the study. Two patterns of missing data can be observed for the response: (1) dropout, when a subject leaves the study prematurely for reasons beyond the control of the investigator, leading to a monotone pattern of nonresponse, or (2) intermittent nonresponse, in which a subject returns to the study after some occasions of nonresponse. Covariates may also be missing, leading to limitations in data analysis. In the presence of missing data, three issues are of main concern: (1) potential serious bias due to systematic differences between the observed data and the missing data, (2) complications in data handling and statistical inferences, and  (3) loss of efficiency. Therefore, in order to make valid inferences it is fundamental to know the missing data mechanism generating the nonresponse and how to handle it.

\cite{little1987} provided a formal framework for dealing with missing data by defining the commonly adopted taxonomy of missing data mechanisms. A nonresponse process is said to be \emph{missing completely at random} (MCAR) if missingness is independent of both unobserved and observed data, and \emph{missing at random} (MAR) if, conditionally on the observed data, the missingness is independent of the unobserved data. When the nonresponse process depends on unobserved quantities it is said to be \emph{missing not at random} (MNAR).  

When data are incomplete, GEE suffers from its frequentist nature and is, in its basic form, valid only under MCAR \citep{liang1986}. The first effort to make GEE applicable to the more realistic MAR scenario was Multiple Imputation (MIGEE), proposed by \cite{little1987}, in which the missing portions of data are multiply imputed taking into account the uncertainty associated with the predicted values. The completed data sets are analyzed by standard methods for complete data, and estimates are combined into a final analysis. Multiple imputation is detailed in the books by \cite{schafer1997}, \cite{little2002} and \cite{carpenter2013}. Later, \cite{robins1995} proposed the Weighted Generalized Estimating Equations (WGEE), which consists in weighting each observation by the inverse of the probability of the data being observed. This method produces consistent estimates provided the weight model is correctly specified.

Doubly robust estimators (DRGEE) arise as a third generalization of ordinary GEE to deal with data subject to MAR mechanism. DR methods have received increasingly attention in the literature in the last decade (see \cite{carpenter2006}, \cite{bang2005}, \cite{tsiatis2006}, \cite{seaman2009}, \cite{chen2011}). The main idea is to supplement the WGEE with a predictive model for the missing quantities conditional on the observed ones. Doubly Robust method requires only the dropout or the conditional model to be correctly specified in order to provide consistent estimates. In the analysis of longitudinal binary data, DR methods have been applied by \cite{seaman2009}, \cite{birhanu2011}, for missing responses and by \cite{chen2011} for intermittently missing response and a single missing covariate. 

Literature of GEE for missing data is comparatively scarce for longitudinal ordinal response. In \cite{toledano1999}, the authors used a weighted GEE method to accommodate arbitrary patterns of a MCAR missing response and missingness in a key covariate subject to a MAR mechanism. A recent paper from \cite{donneau2014a} compared through a simulation study two multiple imputation methods (multivariate normal imputation and ordinal imputation regression) for longitudinal ordinal data subject to dropout. In another paper the same authors compared joint modeling and fully conditional specification approaches for non-monotone missingness \citep{donneau2014b}. The above mentioned papers used single robust versions of GEE and they have treated only a missing MAR baseline covariate or missing MAR response. Thus the use of a DR GEE method for ordinal data with simultaneously intermittently missing response and missing covariate has been in need of further development.

This work was motivated by the Analgesia in Childbirth study which was conducted in Minas Gerais state, Brazil. The main objective of that study was to compare two techniques of analgesia for labor pain in $49$ patients. The response, pain intensity, was subjectively assessed by each patient, and various clinical covariates were taken until delivery. Response and a particular covariate (consumption of oxytocin) were missing for some patients and the MAR mechanism seems to be a reasonable assumption for this data.

In the current paper, it is proposed a doubly robust approach for the analysis of longitudinal ordinal data with intermittently missing response and covariate that is MAR. The proposed methodology, to our knowledge new to the GEE literature of missing ordinal data, can be used for handling arbitrary patterns of missing data in the ordinal response and missingness in a key covariate, as those frequently arising in medical studies. 

The paper is organized as follows. In Section \ref{um_geecomp_mech} are defined the notation for GEE with fully observed data and missing data mechanisms. Section \ref{um_wgeemigee} outlines WGEE and MIGEE approaches. The proposed methodology is established in Section \ref{um_drgee}. A simulation study is presented in Section \ref{um_simula}, in which the finite-sample biases and standard errors are compared for the standard GEE, MIGEE, WGEE and doubly robust versions. Data arising from the Analgesia en Childbirth study are analyzed in Section \ref{um_realdata}. Paper ends with a discussion and future directions in Section \ref{um_discuss}.

\section{GEE for Complete Data and Missing Data Assumptions}\label{um_geecomp_mech}

In this section it is introduced generalized estimating equations for the analysis of fully observed ordinal data. Section \ref{um_geecomp} establishes the model and notation for longitudinal ordinal data. Section \ref{um_mech} presents a series of assumptions related to mechanism causing data to be missing and necessary to be considered in order to build valid estimators.

\subsection{GEE for Longitudinal Ordinal Response}\label{um_geecomp}

Let $O_{it} \in \left\{1,2,\ldots,J \right\}$ be the ordinal response for subject $i \ (i=1,\ldots,n)$ at time $t \ (t=1,\ldots,T_i, \ T_i\leq T)$. As the response has $J$ levels it can be defined $Y_{itj}=I(O_{it}= j)$ for $j=1,\ldots,J$, where $I(A)$ denotes the indicator function. $Y_{itj}$ is converted into the equivalent $(J-1)$-variate vector $\boldsymbol{Y}_{it}=(Y_{it1},\ldots,Y_{it(J-1)})^T$ and let $\boldsymbol{Y}_i=(Y_{i1}^T,\ldots,Y_{iT_i}^T)^T$ the stacked response vector. When $J=2$ the response is binary and $\boldsymbol{Y}_{it}$ is a scalar. Let $\boldsymbol{X}_i=(X_{i1}^T,\ldots,X_{iT_i}^T)^T$ denotes the $T_i\times 1$ covariate vector that may be missing for the $i$-th subject, and  $\boldsymbol{Z}_i=(\boldsymbol{Z}_{i1}^T,\ldots,\boldsymbol{Z}_{iT_i}^T)^T$ the $T_i\times q$ matrix of explanatory variables that are always observed. 

The marginal distribution of $\boldsymbol{Y}_{it}$ is assumed to be multinomial (with sample size $\sum_{j=1}^J Y_{itj}=1$), that is
\begin{equation}
	f(\boldsymbol{Y}_{it}|\boldsymbol{X}_{it},\boldsymbol{Z}_{it},\boldsymbol{\beta})=\prod_{j=1}^J \mu_{itj}^{y_{itj}},
\end{equation}
where $\mu_{itj}=\mu_{itj}(\boldsymbol{\beta})=E(Y_{itj}|\boldsymbol{X}_i,\boldsymbol{Z}_i,\boldsymbol{\beta})=Pr(O_{it}= j|\boldsymbol{X}_i,\boldsymbol{Z}_i,\boldsymbol{\beta})$, is the probability of response $j$ at time $t$ and $\boldsymbol{\beta}$ is a $p \times 1$ vector of parameters. Two common choices for modeling $\mu_{itj}$ are the cumulative logit and probit models. In this work it is assumed a cumulative logit link, that is,
\begin{equation}\label{um_mediaor}
	\mbox{logit} \left[Pr(O_{it}\leq j|X_{it},Z_{it})\right]=\beta_{0j}+X_{it}\beta_x+\boldsymbol{Z}_{it}^T\boldsymbol{\beta}_z ,\ \ \ j=1,\ldots,J-1.
\end{equation}

Formulation in (\ref{um_mediaor}) implies a proportional odds model \citep{mccullagh1980}. In such model the interpretation of $\boldsymbol{\beta}$ is the same regardless of the number of categories (i.e., it is invariant to combination of categories). A desired feature is that the exponential of the parameters is interpreted as an odds ratio \citep{agresti2013}.

Main interest is to make inferences related to the regression parameters \\
$\boldsymbol{\beta}=(\beta_{01},\ldots,\beta_{0,J-1},\beta_x,\boldsymbol{\beta}_z^T)^T$ associated with the $(J-1)\times 1$ marginal probability vectors
\begin{equation*}
	E(Y_{it}|\boldsymbol{X}_i,\boldsymbol{Z}_i)=\boldsymbol{\mu}_{it}(\boldsymbol{\beta})=(\mu_{it1},\ldots,\mu_{it(J-1)})^T.
\end{equation*}
$\boldsymbol{\mu}_{it}$ is grouped to form a vector $E(\boldsymbol{Y}_{i}|\boldsymbol{X}_i,\boldsymbol{Z}_i)=\boldsymbol{\mu}_i=(\boldsymbol{\mu}_{i1}^T,\ldots,\boldsymbol{\mu}_{iT_i}^T)^T$ with the same dimension of $\boldsymbol{Y}_i$.

In order to estimate $\boldsymbol{\beta}$ generalized estimation equations are used (\cite{liang1986}; \cite{lipsitz1994}), which takes the form
\begin{equation}\label{um_gee}
	\boldsymbol{U}(\boldsymbol{\beta})=\sum_{i=1}^n \boldsymbol{U}_i(\boldsymbol{\beta})=\sum_{i=1}^n \boldsymbol{D}_i\boldsymbol{V}_i^{-1}(\boldsymbol{Y}_i-\boldsymbol{\mu}_i)=\boldsymbol{0},
\end{equation}
where $\boldsymbol{D}_i=\frac{\partial \boldsymbol{\mu}_i}{\partial \boldsymbol{\beta}^T}$ and $\boldsymbol{V}_i=\boldsymbol{V}_i(\boldsymbol{\beta},\boldsymbol{\alpha})$ is a $T_i(J-1)\times T_i(J-1)$ ``working covariance'' matrix usually decomposed into the form $\boldsymbol{V}_i(\boldsymbol{\beta},\boldsymbol{\alpha})=\boldsymbol{F}_i^{1/2}(\boldsymbol{\beta})\boldsymbol{C}_i(\boldsymbol{\alpha})\boldsymbol{F}_i^{1/2}(\boldsymbol{\beta})$, where $\boldsymbol{F}_i$ is a matrix containing the marginal variances,  $\boldsymbol{F}_{it}$, given by
\begin{equation*}
	\boldsymbol{F}_{it}=\mbox{diag} \left[\mu_{it1}(1-\mu_{it1}),\ldots,\mu_{it,J-1}(1-\mu_{it,J-1})\right],
\end{equation*}
and $\boldsymbol{C}_i$ is equal to the marginal correlation matrix. The $(J-1)\times (J-1)$ diagonal blocks of $\boldsymbol{C}_i (\boldsymbol{\alpha})$ are $\boldsymbol{F}_{it}^{-1/2}\boldsymbol{V}_{it}\boldsymbol{F}_{it}^{-1/2}$, with $\boldsymbol{V}_{it}=\mbox{diag}(\boldsymbol{\mu}_{it})-\boldsymbol{\mu}_{it}\boldsymbol{\mu}_{it}^T$; and the $(J-1)\times (J-1)$ off-diagonal blocks of $\boldsymbol{C}_i (\boldsymbol{\alpha})$ are $\boldsymbol{\alpha}_{itt'}$, which represents the correlation between $\boldsymbol{Y}_{it}$ and $\boldsymbol{Y}_{it'}$, $t\neq t'$ \citep{lipsitz1994}.

Under mild regularity conditions and correct specification of the marginal mean model in $(\ref{um_mediaor})$, Liang and Zeger (1986) proved that the estimator $\boldsymbol{\hat{\beta}}$, obtained by solving $(\ref{um_gee})$, is consistent and $\sqrt{n}(\boldsymbol{\hat{\beta}}-\boldsymbol{\beta})$ converges in distribution to a $p$-variate normal distribution with mean $\boldsymbol{0}$ and covariance matrix
\begin{equation}\label{um_covbeta}
	\boldsymbol{\mbox{V}}_{\boldsymbol{\beta}}=\lim_{n\rightarrow \infty}n\boldsymbol{\Sigma}_0^{-1}\boldsymbol{\Sigma}_1\boldsymbol{\Sigma}_0^{-1},
\end{equation}
where $\boldsymbol{\Sigma}_0=\sum_{i=1}^n \boldsymbol{D}_i\boldsymbol{V}_i^{-1}\boldsymbol{D}_i^T$, and $\boldsymbol{\Sigma}_1=\sum_{i=1}^n\boldsymbol{D}_i\boldsymbol{V}_i^{-1}\mbox{Cov}(\boldsymbol{Y}_i)\boldsymbol{V}_i^{-1}\boldsymbol{D}_i^T$. In practice, the ``sandwich'' covariance matrix $\boldsymbol{\mbox{V}}_{\boldsymbol{\beta}}$ in $(\ref{um_covbeta})$ is calculated by ignoring the limit and replacing $(\boldsymbol{\beta},\boldsymbol{\alpha})$ and $\mbox{Cov}(\boldsymbol{Y}_i)$ by $(\boldsymbol{\hat{\beta}},\boldsymbol{\hat{\alpha}})$ and $(\boldsymbol{Y}_i-\boldsymbol{\hat{\mu}}_i)(\boldsymbol{Y}_i-\boldsymbol{\hat{\mu}}_i)^T$, respectively \citep{touloumis2013}.

\subsection{Missing Data Framework}\label{um_mech}

For each occasion $t$ it can be defined $R_{it}=0$ if $O_{it}$ and $X_{it}$ are missing, $R_{it}=1$ if $O_{it}$ is missing and $X_{it}$ is observed, $R_{it}=2$ if $O_{it}$ is observed and $X_{it}$ is missing, and $R_{it}=3$ if $O_{it}$ and $X_{it}$ are both observed. Let $\boldsymbol{R}_i=(R_{i1},\ldots,R_{iT_i})^T$, and $\boldsymbol{\bar{R}}_{it}=(R_{i1},\ldots,R_{i,t-1})$.

By specifying conditional models of the form $Pr(R_{it}=r_{it}|\boldsymbol{\bar{R}}_{it},\boldsymbol{O}_i,\boldsymbol{X}_i,\boldsymbol{Z}_i)$ it can be obtained $Pr(\boldsymbol{R}_i=\boldsymbol{r}_i|\boldsymbol{O}_i,\boldsymbol{X}_i,\boldsymbol{Z}_i)$ through $\prod_{i=2}^{T_i}Pr(R_{it}=r_{it}|\boldsymbol{\bar{R}}_{it},\boldsymbol{O}_i,\boldsymbol{X}_i,\boldsymbol{Z}_i)Pr(R_{i1}=r_{i1}|\boldsymbol{O}_i,\boldsymbol{X}_i,\boldsymbol{Z}_i)$. Let $\lambda_{itk}=Pr(R_{it}=k|\boldsymbol{\bar{R}}_{it},\boldsymbol{O}_i,\boldsymbol{X}_i,\boldsymbol{Z}_i)$, for $k=0,1,2,3$. This general formulation encompasses MCAR, MAR and MNAR mechanisms. In particular, the MAR mechanism requires 
\begin{equation}
	Pr(\boldsymbol{R}_{i}=\boldsymbol{r}_{i}|\boldsymbol{O}_i,\boldsymbol{X}_i,\boldsymbol{Z}_i)=
	Pr(\boldsymbol{R}_{i}=\boldsymbol{r}_{i}|\boldsymbol{O}_{i}^o,\boldsymbol{X}_{i}^o,\boldsymbol{Z}_{it}),
\end{equation}
where $\boldsymbol{O}_{i}^o$ and $\boldsymbol{X}_{i}^o$ denotes the observed components of $\boldsymbol{O}_{i}$ and $\boldsymbol{X}_{i}$, respectively. Because $\boldsymbol{R}_i$ is modeled through a product of conditional models, it is natural to make the following further assumption \citep{chen2011}
\begin{equation}
	Pr(R_{it}=r_{it}|\boldsymbol{\bar{R}}_{it},\boldsymbol{O}_i,\boldsymbol{X}_i,\boldsymbol{Z}_i)=
	Pr(R_{it}=r_{it}|\boldsymbol{\bar{R}}_{it},\boldsymbol{\bar{O}}_{it}^o,\boldsymbol{\bar{X}}_{it}^o,\boldsymbol{\bar{Z}}_{it}),
\end{equation}
for each time $t$, where $\boldsymbol{\bar{O}}_{it}^o$ and $\boldsymbol{\bar{X}}_{it}^o$ are the histories of observed responses and covariates up to time $t-1$. 

Let $\pi_{it}=Pr(R_{it}=3|\boldsymbol{O}_i,\boldsymbol{X}_i,\boldsymbol{Z}_i)$ be the marginal probability of observing both $\boldsymbol{O}_i$ and $\boldsymbol{X}_i$ at time $t$, given the entire vectors of responses and covariates. Then, $\pi_{it}$ is expressed by
\begin{equation*}
	\pi_{it}=\sum_{r_{i1},\ldots,r_{i,t-1}}Pr(R_{it}=3,R_{i,t-1}=r_{i,t-1},\ldots,R_{i1}=r_{i1}|\boldsymbol{O}_i,\boldsymbol{X}_i,\boldsymbol{Z}_i).
\end{equation*}
This marginal probability can be expressed in terms of the conditional probabilities $\lambda_{itk}$'s.
Throughout the paper it is required the so-called \emph{positivity assumption}, that is, $\pi_{it}$ must be bounded away from zero. This condition is needed in order to guarantee the existence of $\sqrt{n}$-consistent estimators of $\boldsymbol{\beta}$ \citep{robins1995}.

\section{Available Approaches for Missing Data}\label{um_wgeemigee}

Multiple imputation and weighted generalized estimation equations are two commonly used methods available for missing data under MAR mechanism. These methods are presented in Sections \ref{um_migee} and \ref{um_wgee}, respectively. They serve as the basis for the construction of the doubly robust estimator, presented in Section \ref{um_drgee}.

\subsection{Multiple Imputation Generalized Estimating Equations}\label{um_migee}

%Multiple imputation procedure was formally introduced by \cite{rubin1978} and it has become a popular approach for dealing with the statistical analysis of incomplete data. The key idea is to replace each missing value with a set of $M$ plausible values drawn from the conditional distribution of the unobserved values, given the observed ones. Then a standard GEE is used to analyze each of theses $M$ complete data sets and results from these analyses are then pooled into a single inference using Rubin's rules \citep{rubin1987m}. 

A imputation model commonly used to handle intermittently missing response and covariate, is imputation using chained equations (\cite{vanBuuen1999}, \cite{van2007}), which is more commonly referred to as full conditional specification (FCS). 
This approach specifies conditional distributions for each incomplete variable, conditional on all others variables in the imputation model. Starting from an initial imputation, FCS draws imputations by iterating over the conditional densities.

Denote by $\boldsymbol{\tilde{\beta}}_m$ and $\boldsymbol{\tilde{U}}_m$, respectively, the estimate of $\boldsymbol{\beta}$ and its covariance matrix from the GEE analysis of the $m$-th completed data set, $(m=1,\ldots,M)$. Following \cite{rubin1987m}, the combined point estimate for the parameter of interest $\boldsymbol{\beta}$ from the MI is simply the average of the M complete-data point estimates 
$$
\boldsymbol{\hat{\beta}}_{MI}=\frac{1}{M}\sum_{m=1}^M \boldsymbol{\tilde{\beta}}_m, 
$$
and an estimate of the covariance matrix of $\boldsymbol{\hat{\beta}}_{MI}$ is given by
$$
\boldsymbol{\widehat{U}}_{MI}=\boldsymbol{\tilde{W}}+\left(\frac{M+1}{M}\right)\boldsymbol{\tilde{B}},
$$
where
$$
\boldsymbol{\tilde{W}}=\frac{1}{M}\sum_{m=1}^M \boldsymbol{\tilde{U}}_m \ \ \ \mbox{and} \ \ \
\boldsymbol{\tilde{B}}=\frac{1}{M-1}
\sum_{m=1}^M (\boldsymbol{\tilde{\beta}}_m-\boldsymbol{\hat{\beta}_{MI}})(\boldsymbol{\tilde{\beta}}_m-\boldsymbol{\hat{\beta}_{MI}})'.
$$

\subsection{Weighted Generalized Estimating Equations}\label{um_wgee}
\citet{robins1995} proposed a class of weighted estimating equations to allow for MAR mechanism. In binary longitudinal data, \cite{chen2011} extended the method to accommodate arbitrary patterns of missing response and missing covariate. Their method was adapted here for longitudinal ordinal responses. 

Define a weight matrix $\boldsymbol{\Delta}_i=\left[\delta_{itt'} \right]_{T_i(J_i-1)\times T_i(J_i-1)},\ t=1,\ldots, T_i, t'=1,\ldots, T_i,$ where $\delta_{itt'}=\left\{I(R_{it}=1,R_{it'}=3)+I(R_{it}=3,R_{it'}=3) \right\}/\pi_{itt'}$ for $t\neq t'$, $\delta_{itt}=I(R_{it}=3)/\pi_{it}$, and $\pi_{itt'}=Pr(R_{it}=1,R_{it'}=3|\boldsymbol{O}_i,\boldsymbol{X}_i,\boldsymbol{Z}_i)+Pr(R_{it}=3,R_{it'}=3|\boldsymbol{O}_i,\boldsymbol{X}_i,\boldsymbol{Z}_i)$. Let $\boldsymbol{M}_i=\boldsymbol{F}_i^{-1/2}(\boldsymbol{C}_i^{-1}\boldsymbol{\cdot}\boldsymbol{\Delta}_i)\boldsymbol{F}_i^{-1/2}$ where $\boldsymbol{A\cdot B}=\left[a_{it}\cdot b_{it} \right]$ denotes the Hadamard product of matrix $\boldsymbol{A}=\left[a_{it} \right]$ and $\boldsymbol{B}=\left[b_{it} \right]$. 

The weighted generalized estimating equations (WGEE) for $\boldsymbol{\beta}$ are given by
\begin{equation}\label{wgee2}
	\boldsymbol{U}(\boldsymbol{\beta},\boldsymbol{\psi})=\sum_{i=1}^n \boldsymbol{U}_i(\boldsymbol{\beta},\boldsymbol{\psi})=\boldsymbol{0},
\end{equation}
where $\boldsymbol{U}_i(\boldsymbol{\beta},\boldsymbol{\psi})=\boldsymbol{D}_i\boldsymbol{M}_i(\boldsymbol{Y}_i-\boldsymbol{\mu}_i)$. A consistent estimate for $\boldsymbol{\beta}$ can be obtained by solving (\ref{wgee2}), under the correct specification of the missing data model.

To model $\lambda_{itk}$ it is adopted a politomic logistic regression, with $\lambda_{it0}$ as the reference category, that is
\begin{equation}
	\log\left( \frac{\lambda_{itk}}{\lambda_{it0}}\right)=\boldsymbol{u}_{itk}^T\psi_k, \ \ \ k=1,2,3,
\end{equation}
where the covariates $\boldsymbol{u}_{itk}$ are some function of $\left\{\boldsymbol{\bar{R}}_{it},\boldsymbol{\bar{O}}_{it}^o,\boldsymbol{\bar{X}}_{it}^o,\boldsymbol{\bar{Z}}_{it}\right\}$. 
%Let $\boldsymbol{\psi}=(\psi_1^T,\psi_2^T,\psi_3^T)^T$.

\section{Doubly Robust GEE for Longitudinal Ordinal Data}\label{um_drgee}

Some authors (e.g., \cite{scharfstein1999}, \cite{tsiatis2006}) noted that adding a term of expectation zero, say $\phi(\cdot)$, to the inverse probability weighted estimators would still result in consistent estimates under a MAR mechanism. The solutions of these augmented estimating equations give rise to the so-called \emph{doubly robust} estimators. 

\cite{chen2011} showed that the optimal $\phi_{opt}$ for missing response and covariate is given by $\phi_{opt}=E_{(\boldsymbol{Y}_i^m,\boldsymbol{X}_i^m|\boldsymbol{Y}_i^o,\boldsymbol{X}_i^o,\boldsymbol{Z}_i,\boldsymbol{R}_i)}\left\{\boldsymbol{D}_i\boldsymbol{N}_i(\boldsymbol{Y}_i-\boldsymbol{\mu}_i)\right\}$, with $\boldsymbol{N}_i=
\boldsymbol{F}_i^{-1/2}\left\{\boldsymbol{C}_i^{-1}\boldsymbol{\cdot}(\boldsymbol{11}^T-\boldsymbol{\Delta}_i)\right\}\boldsymbol{F}_i^{-1/2}$, where $\boldsymbol{1}$ is a vector of 1's of length $T_i(J-1)$, and $\boldsymbol{Y}_i^m$ and $\boldsymbol{X}_i^m$ denote the missing components of $\boldsymbol{Y}_i$ and $\boldsymbol{X}_i$, respectively.

An improved estimate for $\boldsymbol{\beta}$ can then be obtained by solving the estimating equations
\begin{equation}\label{dr}
	%\scriptsize
	\boldsymbol{S}_1(\boldsymbol{\theta})=\sum_{i=1}^n \boldsymbol{S}_{1i}(\boldsymbol{\theta})=\sum_{i=1}^n \left[
	\boldsymbol{D}_i\boldsymbol{M}_i(\boldsymbol{Y}_i-\boldsymbol{\mu}_i)+
	E_{(\boldsymbol{Y}_i^m,\boldsymbol{X}_i^m|\boldsymbol{Y}_i^o,\boldsymbol{X}_i^o,\boldsymbol{Z}_i,\boldsymbol{R}_i)}\left\{\boldsymbol{D}_i\boldsymbol{N}_i(\boldsymbol{Y}_i-\boldsymbol{\mu}_i)\right\}
	\right]=\boldsymbol{0}.
\end{equation}
\normalsize
The estimator for $\boldsymbol{\beta}$ in (\ref{dr}) is doubly-robust in the sense that it is consistent if \emph{at least one} of the missing data model or the covariate model is correctly specified.

Applications of doubly robust estimators in longitudinal settings include \cite{bang2005}, \cite{seaman2009}, \cite{chen2011} and \cite{birhanu2011}. Those developments focus mainly on binary response and have not been, to our knowledge, investigated with ordinal longitudinal data. In this work it is considered a longitudinal response measured on a ordinal scale. 

The referred expectation in the second part of (\ref{dr}) is over the conditional distribution of $(\boldsymbol{Y}_i^m,\boldsymbol{X}_i^m|\boldsymbol{Y}_i^o,\boldsymbol{X}_i^o,\boldsymbol{Z}_i,\boldsymbol{R}_i)$, which can be written as
\begin{eqnarray*}
	P(\boldsymbol{Y}_i^m=\boldsymbol{y}_i^m,\boldsymbol{X}_i^m=\boldsymbol{x}_i^m|\boldsymbol{Y}_i^o,\boldsymbol{X}_i^o,\boldsymbol{Z}_i,\boldsymbol{R}_i;\boldsymbol{\beta}^*,\boldsymbol{\gamma})
	&=&P(\boldsymbol{Y}_i^m=\boldsymbol{y}_i^m,\boldsymbol{X}_i^m=\boldsymbol{x}_i^m|\boldsymbol{Y}_i^o,\boldsymbol{X}_i^o,\boldsymbol{Z}_i;\boldsymbol{\beta}^*,\boldsymbol{\gamma})\\
	&=&P(\boldsymbol{Y}_i^m=\boldsymbol{y}_i^m|\boldsymbol{Y}_i^o,\boldsymbol{X}_i=\boldsymbol{x}_i,\boldsymbol{Z}_i;\boldsymbol{\beta}^*)\\
	&&\times
	P(\boldsymbol{X}_i^m=\boldsymbol{x}_i^m|\boldsymbol{Y}_i^o,\boldsymbol{X}_i^o,\boldsymbol{Z}_i;\boldsymbol{\gamma}).
\end{eqnarray*}

The multivariate distribution $P(\boldsymbol{Y}_i^m=\boldsymbol{y}_i^m|\boldsymbol{Y}_i^o,\boldsymbol{X}_i=\boldsymbol{x}_i,\boldsymbol{Z}_i;\boldsymbol{\beta}^*)$ is expressed through a product of univariate ordinal models. 
%Discrete and continuous cases of the missing covariate are treated in the Sections 4.1 and 4.2.

\subsection{Estimation for the Nuisance Parameters}

The method of maximum likelihood is employed to estimate $\boldsymbol{\psi}$. The log likelihood of the politomic logistic model for $\boldsymbol{\psi}$ has the form
\begin{equation*}
	l(\boldsymbol{\boldsymbol{\psi}})=\sum_{i=1}^n l_i(\boldsymbol{\psi})=\sum_{i=1}^n\sum_{t=1}^{T_i}\sum_{k=0}^3
	I(R_{it}=k)\log(\lambda_{itk}),
\end{equation*}
with corresponding score function given by
\begin{eqnarray*}
	\boldsymbol{S}_2(\boldsymbol{\psi})=\sum_{i=1}^n \boldsymbol{S}_{2i}(\boldsymbol{\psi})=\sum_{i=1}^n\sum_{t=1}^{T_i}\sum_{k=0}^3\frac{I(R_{it}=k)}{\lambda_{itk}}\frac{\partial \lambda_{itk}}{\partial \boldsymbol{\psi}^T}.
\end{eqnarray*}
The maximum likelihood estimator $\boldsymbol{\hat{\psi}}$, is obtained by solving $\boldsymbol{S}_2(\boldsymbol{\psi})=\boldsymbol{0}$.

For the missing covariate model, the observed likelihood function for $\boldsymbol{\gamma}$ is 
\begin{equation*}
	L_3(\boldsymbol{\gamma})=\prod_{i=1}^n \int Pr(\boldsymbol{X}_i|\boldsymbol{Z}_i,\boldsymbol{Y}_i)d\boldsymbol{X}_i^m,
\end{equation*}
with score function $\boldsymbol{S}_3(\boldsymbol{\gamma})=\sum_{i=1}^n \boldsymbol{S}_{3i}(\boldsymbol{\gamma})$, where $\boldsymbol{S}_{3i}(\boldsymbol{\gamma})=\partial \log \int Pr(\boldsymbol{X}_i|\boldsymbol{Z}_i,\boldsymbol{Y}_i))d\boldsymbol{X}_i^m
/\partial \boldsymbol{\gamma}^T$. Similarly, a consistent estimator of $\boldsymbol{\gamma}$ can be obtained by solving $\boldsymbol{S}_3(\boldsymbol{\gamma})=\boldsymbol{0}$.

\subsection{Estimation and Inference for the Doubly Robust Method}

Let's denote the vector of all parameters as $\boldsymbol{\theta}=(\boldsymbol{\beta}^T,\boldsymbol{\psi}^T,\boldsymbol{\gamma}^T)^T$. Our primary interest lies is in estimating $\boldsymbol{\beta}$. Such task can be accomplished by plugging in the estimates $\boldsymbol{\hat{\psi}}$ and $\boldsymbol{\hat{\gamma}}$ in (\ref{dr}) and solving the estimating equations for $\boldsymbol{\beta}$, that is,
\begin{equation}\label{dr2}
	\boldsymbol{S}_1(\boldsymbol{\beta},\boldsymbol{\hat{\psi}},\boldsymbol{\hat{\gamma}})=\sum_{i=1}^n\boldsymbol{S}_{1i}(\boldsymbol{\beta},\boldsymbol{\hat{\psi}},\boldsymbol{\hat{\gamma}})=0.
\end{equation}

Second term of $\boldsymbol{S}_{1i}$ can be written, for $\boldsymbol{X}$ discrete, as
\begin{equation*}
	E_{(\boldsymbol{Y}_i^m,\boldsymbol{X}_i^m|\boldsymbol{Y}_i^o,\boldsymbol{X}_i^o,\boldsymbol{Z}_i,\boldsymbol{R}_i)}\left\{\boldsymbol{D}_i\boldsymbol{N}_i(\boldsymbol{Y}_i-\boldsymbol{\mu}_i)\right\}=
	\sum_{(\boldsymbol{y}_i^m,\boldsymbol{x}_i^m)}w_{ixy}
	\left\{\boldsymbol{D}_i\boldsymbol{N}_i(\boldsymbol{Y}_i-\boldsymbol{\mu}_i)\right\},
\end{equation*}
where the weight $w_{ixy}$ is given by
\begin{eqnarray*}
	w_{ixy}&=&P(\boldsymbol{Y}_i^m=\boldsymbol{y}_i^m|\boldsymbol{Y}_i^o,\boldsymbol{X}_i=\boldsymbol{x}_i,\boldsymbol{Z}_i;\boldsymbol{\beta}^*)\times
	P(\boldsymbol{X}_i^m=\boldsymbol{x}_i^m|\boldsymbol{Y}_i^o,\boldsymbol{X}_i^o,\boldsymbol{Z}_i;\boldsymbol{\hat{\gamma}}).
\end{eqnarray*}

In the case of $\boldsymbol{X}$ continuous, the second term in $\boldsymbol{S}_{1i}$ takes the form\\
\begin{equation*}
	E_{(\boldsymbol{Y}_i^m,\boldsymbol{X}_i^m|\boldsymbol{Y}_i^o,\boldsymbol{X}_i^o,\boldsymbol{Z}_i,\boldsymbol{R}_i)}\left\{\boldsymbol{D}_i\boldsymbol{N}_i(\boldsymbol{Y}_i-\boldsymbol{\mu}_i)\right\}=
	\int_{(\boldsymbol{Y}_i^m,\boldsymbol{X}_i^m)}w_{ixy}
	\left\{\boldsymbol{D}_i\boldsymbol{N}_i(\boldsymbol{Y}_i-\boldsymbol{\mu}_i)\right\}d\boldsymbol{Y}_i^m\boldsymbol{X}_i^m,
\end{equation*}
with conditional probability
\begin{eqnarray*}
	w_{ixy}&=&P(\boldsymbol{Y}_i^m=\boldsymbol{y}_i^m|\boldsymbol{Y}_i^o,\boldsymbol{X}_i=\boldsymbol{x}_i,\boldsymbol{Z}_i;\boldsymbol{\beta}^*)\times
	P(\boldsymbol{X}_i^m=\boldsymbol{x}_i^m|\boldsymbol{Y}_i^o,\boldsymbol{X}_i^o,\boldsymbol{Z}_i;\boldsymbol{\hat{\gamma}}).
\end{eqnarray*}

The expectation in (\ref{dr}) can be cumbersome, depending on the missing data pattern. In such case, instead of using numerical integration, a Monte Carlo method can be applied to approximate the corresponding integral. 

In this work it is assumed independence working correlation and it is adopted a sandwich standard error as given in Appendix.

\section{Simulation Study}\label{um_simula}

A small simulation study taking into account different sample sizes was conducted in order to quantify the bias and precision under misspecification of the predictive models.  It is considered a study with $T_i=T=3$ repeated ordinal measures (with three categories) and two covariates (quantitative and qualitative). The true marginal model is 
\begin{equation}\label{um_gera.resp}
	logit \ Pr(O_{it}\leq j|X_{it},Z_{it})=\beta_{0j}+\beta_1X_{it}+\beta_2Z_{it}, \ \ \ j=1,2.
\end{equation} 
where $Z_{it}$ is normal with unity variance and mean $(0,0.5,1)$ for $t=1,2,3$. 

The binary covariate $X_{it}$ may be missing at some time points and is generated according to
\begin{equation}\label{um_gera.cov}
	logit \ Pr(X_{it}=1|\bar{X}_{it},Z_{it})=\gamma_0+\gamma_1X_{i,t-1}+\gamma_2Z_{it}.
\end{equation}
It is assumed $\beta_{01}=-0.4$, $\beta_{02}=1.2$, $\beta_1=-0.5$, $\beta_2=0.5$, $\gamma_0=log(1)$, $\gamma_1=2$ and $\gamma_2=2$. The correlated ordinal responses were generated according to the algorithm proposed by Touloumis \citep{SimCorMultRes} with constant correlation between the latent vectors set equal to $\rho=0.9$.

As independent estimating equations were fitted, $R_{it}$ can be defined as the indicator of observing both $O_{it}$ and $X_{it}$, and it was taken
\begin{equation}\label{um_gera.perda}
	log\left(\frac{Pr(R_{it}=1)}{Pr(R_{it}=0)} \right)=\psi_{0t}+\psi_1I(R_{i,t-1}=1)+\psi_2O_{i,t-1}^*+\psi_3X_{i,t-1}^*+\psi_4Z_{it}, \ \ t=2,3,
\end{equation}
where $O_{i,t-1}^*=O_{i,t-1}$, if $O_{i,t-1}$ is observed and $0$ otherwise, and $X_{i,t-1}^*=X_{i,t-1}$if $X_{i,t-1}$ is observed and $0$ otherwise. The true values are taken as $\psi_{02}=6.6$, $\psi_{03}=6$, $\psi_{1}=2$, $\psi_{2}=-2$, $\psi_{3}=-2$ and $\psi_{4}=2$. It was observed about $24\%$ of missing observations under this setup.

For comparison purposes, it was considered ordinary GEE for the complete and available data, respectively, weighted GEE (WGEE), multiple imputation (MIGEE) by chained equations \citep{mice} with $M=10$, and the proposed doubly robust version (DRGEE). In order to investigate robustness of these methods, the predicted models were also misspecified by omitting the covariate $X_{t-1}$ from the covariate model (\ref{um_gera.cov}) or the missing data model (\ref{um_gera.perda}).

Results are summarized in Table \ref{um_tab1}. In each of the $S=1000$ Monte Carlo replications it was obtained the relative bias percentage for each parameter, defined as $100\times(\hat{\beta}-\beta)/\beta$, its standard deviation obtained through the sandwich estimator, and the coverage probability as a nominal 95\%.

\begin{table}
	\begin{center}
		\scriptsize
		\caption{Relative bias percentage, standard deviation and empirical coverage for $1000$ simulations of incomplete covariate and response data.}\label{um_tab1}
		\begin{tabular}{rrrrrrrrrrrrrrr}
			\hline
			\hline
			& \multicolumn{4}{c}{Empirical Bias} &       & \multicolumn{4}{c}{Standard Deviation} &       & \multicolumn{4}{c}{Empirical Coverage} \\
			\cline{2-5}\cline{7-10}\cline{12-15}          & $\beta_{01}$ & $\beta_{02}$ & $\beta_1$ & $\beta_2$ & \textbf{} & $\beta_{01}$ & $\beta_{02}$ & $\beta_1$ & $\beta_2$ &       & $\beta_{01}$ & $\beta_{02}$ & $\beta_1$ & $\beta_2$ \\
			\hline
			& \multicolumn{14}{c}{$n=50$} \\
			Complete & -4.60 & 7.89  & 15.07 & 7.38  &       & 0.379 & 0.410 & 0.458 & 0.179 &       & 0.94  & 0.94  & 0.97  & 0.95 \\
			Available & -19.81 & 15.37 & -8.36 & -11.66 &       & 0.386 & 0.420 & 0.473 & 0.190 &       & 0.94  & 0.92  & 0.97  & 0.94 \\
			WGEE($r^+$) & -12.33 & 12.02 & 19.96 & 1.74  &       & 0.406 & 0.440 & 0.513 & 0.204 &       & 0.94  & 0.94  & 0.97  & 0.95 \\
			WGEE($r^-$) & 1.50  & 7.43  & 1.31  & 0.51  &       & 0.413 & 0.450 & 0.508 & 0.201 &       & 0.95  & 0.93  & 0.96  & 0.95 \\
			MIGEE($x^+$) & 22.17 & -3.44 & -38.26 & -4.73 &       & 0.386 & 0.417 & 0.480 & 0.186 &       & 0.96  & 0.96  & 0.97  & 0.96 \\
			MIGEE($x^-$) & 15.81 & -0.13 & -26.18 & -2.54 &       & 0.387 & 0.417 & 0.481 & 0.187 &       & 0.95  & 0.96  & 0.97  & 0.96 \\
			DRGEE($x^+,r^+$) & -8.86 & 9.67  & 22.20 & 8.99  &       & 0.410 & 0.442 & 0.513 & 0.199 &       & 0.95  & 0.94  & 0.97  & 0.95 \\
			DRGEE($x^-,r^+$)  & -7.77 & 9.44  & 19.33 & 7.91  &       & 0.471 & 0.507 & 0.579 & 0.204 &       & 0.95  & 0.94  & 0.97  & 0.95 \\
			DRGEE($x^+,r^-$) & -7.95 & 10.05 & 20.58 & 8.26  &       & 0.433 & 0.473 & 0.540 & 0.197 &       & 0.95  & 0.93  & 0.96  & 0.95 \\
			DRGEE($x^-,r^-$) & 2.86  & 6.42  & -0.28 & 2.99  &       & 0.414 & 0.450 & 0.529 & 0.200 &       & 0.95  & 0.93  & 0.97  & 0.95 \\
			& \multicolumn{14}{c}{$n=150$} \\
			Complete & -1.36 & 1.53  & 5.13  & 1.99  &       & 0.220 & 0.236 & 0.264 & 0.103 &       & 0.94  & 0.94  & 0.96  & 0.95 \\
			Available & -17.50 & 8.94  & -18.41 & -17.15 &       & 0.224 & 0.242 & 0.273 & 0.109 &       & 0.91  & 0.92  & 0.94  & 0.91 \\
			WGEE($r^+$) & -5.84 & 3.41  & 13.47 & 1.36  &       & 0.247 & 0.267 & 0.317 & 0.126 &       & 0.93  & 0.93  & 0.95  & 0.94 \\
			WGEE($r^-$) & 10.20 & -2.04 & -12.91 & -2.59 &       & 0.253 & 0.279 & 0.307 & 0.120 &       & 0.94  & 0.93  & 0.93  & 0.94 \\
			MIGEE($x^+$) & 8.22  & -2.57 & -14.36 & -2.40 &       & 0.228 & 0.244 & 0.283 & 0.108 &       & 0.94  & 0.94  & 0.96  & 0.95 \\
			MIGEE($x^-$) & 9.08  & -1.97 & -16.33 & -3.60 &       & 0.226 & 0.242 & 0.280 & 0.108 &       & 0.93  & 0.93  & 0.95  & 0.94 \\
			DRGEE($x^+,r^+$) & -4.21 & 2.64  & 11.90 & 4.12  &       & 0.249 & 0.269 & 0.317 & 0.117 &       & 0.94  & 0.94  & 0.96  & 0.95 \\
			DRGEE($x^-,r^+$)  & -5.02 & 3.07  & 12.26 & 3.63  &       & 0.320 & 0.345 & 0.396 & 0.119 &       & 0.94  & 0.94  & 0.95  & 0.95 \\
			DRGEE($x^+,r^-$) & -1.86 & 1.99  & 7.26  & 2.95  &       & 0.249 & 0.275 & 0.313 & 0.113 &       & 0.94  & 0.94  & 0.95  & 0.94 \\
			DRGEE($x^-,r^-$) & 9.88  & -1.86 & -15.60 & -2.86 &       & 0.246 & 0.269 & 0.317 & 0.116 &       & 0.94  & 0.93  & 0.94  & 0.94 \\
			& \multicolumn{14}{c}{$n=300$} \\
			Complete & 1.42  & 0.19  & -0.04 & 0.85  &       & 0.156 & 0.167 & 0.187 & 0.072 &       & 0.96  & 0.94  & 0.94  & 0.95 \\
			Available & -14.82 & 7.67  & -23.24 & -18.14 &       & 0.159 & 0.171 & 0.194 & 0.077 &       & 0.93  & 0.91  & 0.92  & 0.87 \\
			WGEE($r^+$) & -0.24 & 0.75  & 5.61  & 1.23  &       & 0.182 & 0.198 & 0.235 & 0.095 &       & 0.95  & 0.94  & 0.94  & 0.93 \\
			WGEE($r^-$) & 15.64 & -4.50 & -22.17 & -3.89 &       & 0.184 & 0.205 & 0.222 & 0.088 &       & 0.95  & 0.94  & 0.92  & 0.94 \\
			MIGEE($x^+$) & 6.23  & -1.86 & -9.95 & -1.44 &       & 0.162 & 0.174 & 0.201 & 0.076 &       & 0.95  & 0.94  & 0.94  & 0.95 \\
			MIGEE($x^-$) & 10.27 & -2.70 & -18.22 & -3.89 &       & 0.160 & 0.172 & 0.200 & 0.076 &       & 0.94  & 0.92  & 0.92  & 0.94 \\
			DRGEE($x^+,r^+$) & -0.06 & 0.83  & 2.82  & 1.28  &       & 0.185 & 0.202 & 0.242 & 0.089 &       & 0.96  & 0.95  & 0.95  & 0.94 \\
			DRGEE($x^-,r^+$)  & -0.71 & 1.08  & 3.84  & 1.43  &       & 0.194 & 0.213 & 0.251 & 0.088 &       & 0.96  & 0.95  & 0.95  & 0.94 \\
			DRGEE($x^+,r^-$) & 2.53  & 0.03  & -2.11 & 0.12  &       & 0.178 & 0.198 & 0.224 & 0.082 &       & 0.96  & 0.96  & 0.94  & 0.94 \\
			DRGEE($x^-,r^-$) & 13.86 & -3.69 & -24.29 & -5.51 &       & 0.176 & 0.194 & 0.228 & 0.084 &       & 0.95  & 0.94  & 0.93  & 0.94 \\
			& \multicolumn{14}{c}{$n=600$} \\
			Complete & -1.01 & 0.57  & 1.93  & 0.82  &       & 0.110 & 0.118 & 0.132 & 0.051 &       & 0.95  & 0.93  & 0.94  & 0.94 \\
			Available & -16.78 & 7.85  & -22.20 & -18.43 &       & 0.112 & 0.121 & 0.137 & 0.054 &       & 0.91  & 0.88  & 0.92  & 0.78 \\
			WGEE($r^+$) & -1.00 & 0.28  & 5.43  & 1.25  &       & 0.132 & 0.144 & 0.170 & 0.070 &       & 0.95  & 0.94  & 0.95  & 0.95 \\
			WGEE($r^-$) & 14.87 & -4.99 & -21.53 & -3.03 &       & 0.133 & 0.149 & 0.158 & 0.064 &       & 0.94  & 0.94  & 0.92  & 0.93 \\
			MIGEE($x^+$) & 1.76  & -0.66 & -3.63 & -0.39 &       & 0.115 & 0.123 & 0.143 & 0.054 &       & 0.95  & 0.94  & 0.95  & 0.94 \\
			MIGEE($x^-$) & 6.96  & -2.01 & -14.18 & -3.38 &       & 0.113 & 0.122 & 0.141 & 0.054 &       & 0.94  & 0.93  & 0.93  & 0.93 \\
			DRGEE($x^+,r^+$) & -1.49 & 0.59  & 3.19  & 0.98  &       & 0.132 & 0.144 & 0.174 & 0.062 &       & 0.96  & 0.94  & 0.94  & 0.94 \\
			DRGEE($x^-,r^+$) & -1.39 & 0.56  & 2.96  & 0.91  &       & 0.136 & 0.149 & 0.178 & 0.062 &       & 0.96  & 0.94  & 0.96  & 0.95 \\
			DRGEE($x^+,r^-$) & 0.42  & 0.08  & -0.12 & 0.47  &       & 0.126 & 0.140 & 0.159 & 0.058 &       & 0.96  & 0.95  & 0.95  & 0.93 \\
			DRGEE($x^-,r^-$) & 12.24 & -3.82 & -23.18 & -5.35 &       & 0.124 & 0.138 & 0.161 & 0.060 &       & 0.93  & 0.93  & 0.92  & 0.93 \\
			\hline
			\multicolumn{14}{c}{``$^+$" indicates correctly specified model and ``$^-$" indicates misspecified model omitting the $X_t$ predictor}
		\end{tabular}
	\end{center}
\end{table}

Specially, MAR missingness impact over the response and the covariate is observed for all the regression parameters, the largest relative bias occur in binary covariate $X$. This comes in addition to the natural increase of parameter uncertainty. Bias in the intercept coefficients imply incorrect predicted probabilities for the levels of the response, whereas bias for parameter estimates associated with the regression covariates may erroneously attenuate or highlight an effect, thus leading to misinterpretations related to the importance of a given predictor on the longitudinal dynamics of the ordinal response.

It can be observed that for small sample size even the GEE with for complete data presents a certain degree of bias. Increasing the sample size allows to clarify the performance distinctions among the compared methods. WGEE and MIGEE methods are valid when the model for the weight or the imputation model, respectively, are correctly specified. In this case it is noted that both methods give good results for large sample sizes, the main distinction between them being due to the greater variability of the estimates for the weighted estimator. 

Doubly robust method requires the simultaneous specification of two predictive model. When at least one of them is correctly specified the resulting estimator is still consistent. Estimates are, on average, closer to those obtained with fully observed data compared to WGEE or MIGEE. This behavior is systematic and it can be observed to all parameters. By increasing the sample size the estimates from DRGEE present empirical bias in general smaller than their single robust competitors. This is specially true for the parameter associated with the incomplete binary variable $X$. Regarding the uncertainty of parameter estimates, it is noted that the variability in DRGEE is greater than multiple imputation, but of the same order as of the weighted method. Further, the efficiency of the doubly robust estimates appears relatively more sensitive to misspecification of the weight model than the covariate model. Empirical coverage rates were acceptable for correctly specified WGEE and MIGEE as well as for DRGEE when at least one of predictive models are correctly specified.

\begin{figure}
	\centering
	\includegraphics[height=9cm,width=16cm]{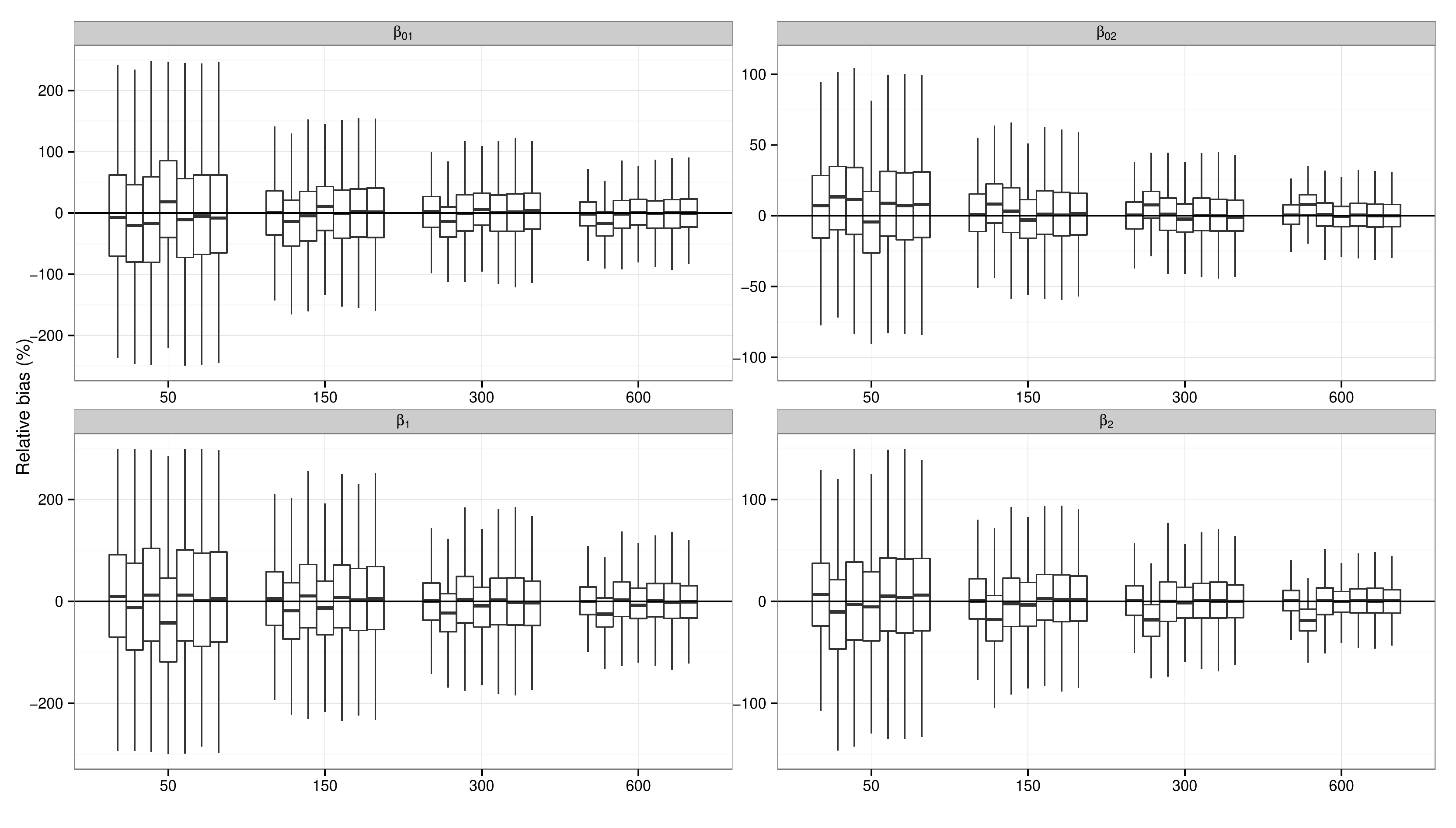}
	\caption{Boxplot of the relative bias for parameter estimates under correctly specified models}
	\label{um_figcor}
\end{figure}

Figure \ref{um_figcor} shows boxplots of the percentage relative bias for the methods expected to be valid. The boxs represent, respectively, GEE for complete and available data, WGEE($r^+$), MIGEE($x^+$), DRGEE($x^+,r^+$), DRGEE($x^-,r^+$) and DRGEE($x^+,r^-$). As the degree of bias is different in the parameter estimates, for easy of visualization, they are represented in different scales. It can be seen that the estimates with larger variability were those associated with the first intercept and the covariate with incomplete values. Proposed method presents median relative bias close to zero and very similar to those observed with complete data for all sample sizes. Variability of the DR estimators is slightly larger than MI but of the same order as the weighted estimator. As expected, for all methods it can be noticed a decrease in the relative bias with increasing sample size, reflecting their theoretical asymptotic consistency.

\begin{figure}
	\centering
	\includegraphics[height=9cm,width=16cm]{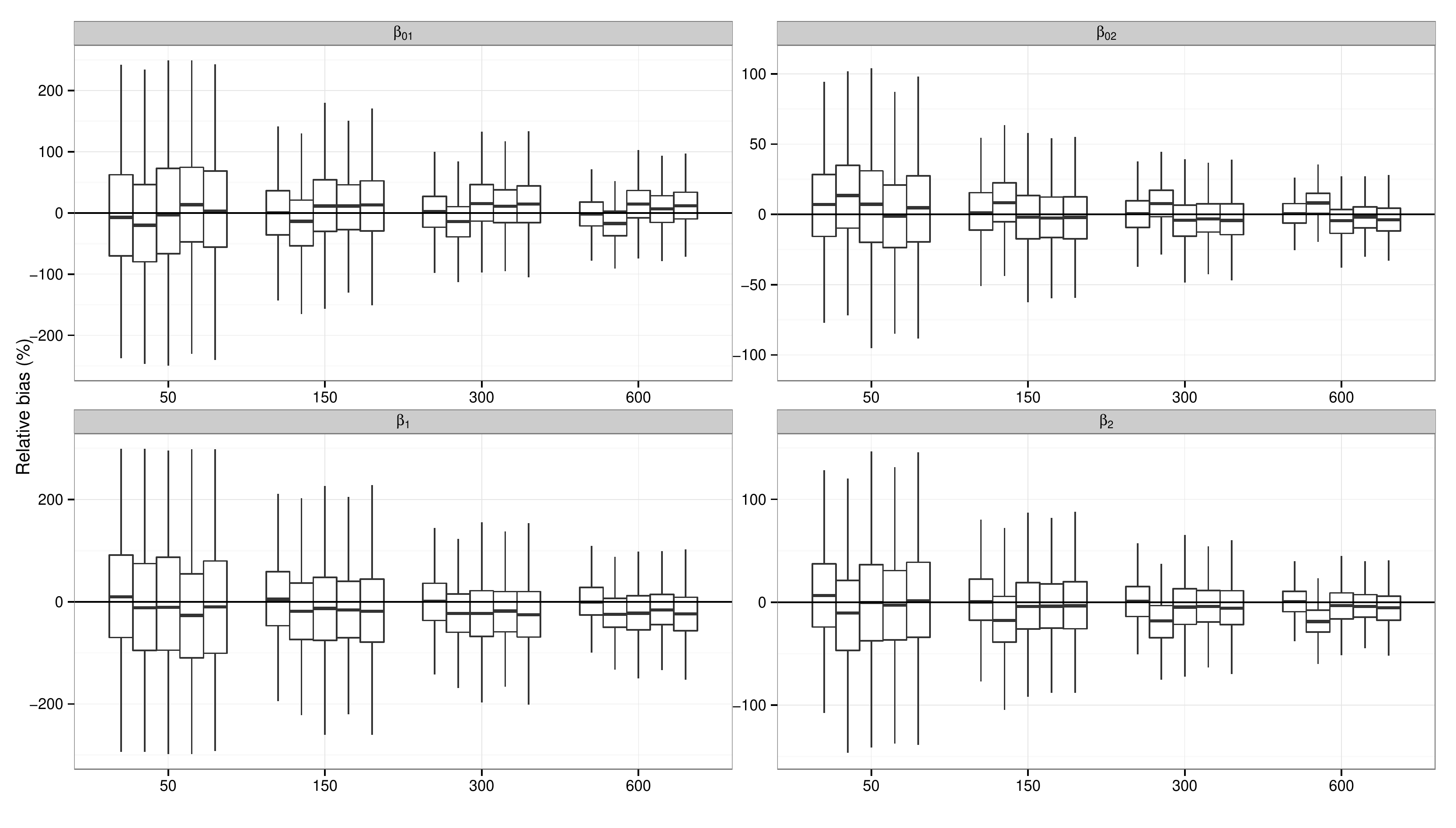}
	\caption{Boxplot of the relative bias for the estimates with incorrectly specified models}
	\label{um_figwro}
\end{figure}

Figure \ref{um_figwro} allows the comparison of methods incorrectly specified. The boxs represent, respectively, GEE for complete and available data, WGEE($r^-$), MIGEE($x^-$), and DRGEE($x^-,r^-$).

When a key covariate is omitted from the weight model and/or the imputation one, it is expected all methods to be biased. This is specially true for the first intercept and for the incomplete binary covariate. Bias for MIGEE($x^-$) seemed to get smaller than WGEE($r^-$) as sample sizes increases. Bias for DRGEE($x^-,r^-$) was comparable to WGEE($r^-$) and slightly higher than multiple imputation for large sample sizes. In terms of variability of the estimates, the same pattern is observed as with the correctly specified models. That is, the multiple imputation is more efficient, followed by the doubly robust estimator and the weighed estimator.

\section{Data Analysis: Analgesia in Childbirth}\label{um_realdata}

This study was conducted in Minas Gerais state, Brazil, in order to compare two techniques of analgesia for labor pain. There were $49$ patients who were monitored during their entire labor period until childbirth. Pain intensity was subjectively assessed by the patient, and measurements of blood pressure, maternal heart rate, consumption of oxytocin, sedation level, signs of respiratory depression, apnea, and other variables were recorded. One of the techniques used was epidural analgesia (the gold standard), which is a local anesthetic. The other one, whose efficiency was to be compared to the gold standard, involved continuous intravenous infusion of remifentanil. 

The response of interest is the intensity of pain as measured by a Visual Analog Scale (VAS) ($1$: tolerable and mild pain; $2$: moderate pain that causes discomfort; $3$: intense and unbearable pain). Three measurements ($0, 60, 90$ minutes) were selected for data analysis. Predictor variables considered were treatment GROUP ($0$: peridural; $1$: remifentanil), AGE (in years), DU (uterine dilatation), and OXYT (consumption of oxytocin). The OXYT is a time-varying ordinal covariate, coded as $1$, if no consumption, $2$, if consumption equals to $10$ or $30$, and $3$, if consumption equals or above $45$. Thes eother covariates were chosen after a previous exploratory analysis.

The response and oxytocin consumption was missing for $9$ patients at the time $60$ and for $18$ patients at time $90$. Missing is due to childbirth happened before $60$ or $90$ minutes. Therefore a MAR mechanism seems to be a reasonable assumption for this data set. The others covariates in the analysis were fully observed.

For the ordinal response it was used the following proportional odds model
\begin{equation}\label{um_mod.resp}
	logit \ Pr(PAIN_{itj}\leq j|\boldsymbol{u_{it}})=\beta_{0j}+\boldsymbol{u}_{it}^T\boldsymbol{\beta}, \ \ \ j=1,2, \ \ t=1,2,3,
\end{equation} 
where $\boldsymbol{u}_{it}$ is the covariate vector at time $t$, and it is formed by TIME, GROUP, AGE, DU and OXIT. In response model, TIME predictor was expressed in hours rather than minutes.

When using WGEE or DRGGE it is necessary to correctly model $\pi_i$ in order to obtain consistent estimates of $\boldsymbol{\beta}$. For the missing data process $R_{it}$ was defined as the indicator of observing both $PAIN_{it}$ and $OXIT_{it}$, and take the following form
\begin{equation}\label{um_mod.perda}
	log\left(\frac{Pr(R_{it}=1)}{Pr(R_{it}=0)} \right)=\psi_{0t}+\boldsymbol{w}_{it}^T\boldsymbol{\psi}, \ \ t=2,3,
\end{equation}
where $\boldsymbol{w}_{it}$ includes GROUP, AGE, DU, histories of OXYT and PAIN, and the previous indicator of missing data.

The distribution of the missing covariate OXYT also needs to be specified in a predictive model. With this aim, it was assumed a proportional odds model of the form
\begin{equation}\label{um_mod.cov}
	logito \ Pr(OXYT_{itj}\leq j|\boldsymbol{v_{it}})=\gamma_{0j}+\boldsymbol{v}_{it}^T\boldsymbol{\gamma}, \ \ \ j=1,2, \ \ t=2,3
\end{equation} 
where $\boldsymbol{v}_{it}$ includes main effects for GROUP, AGE, and DU. It can be observed that the estimate of $\boldsymbol{\gamma}$ is not of interest, however it is necessary to model the missing mechanism related to covariate as close as possible to true in order to obtain valid estimates of $\boldsymbol{\beta}$. The same is true for the missing data process. All predictors in this model process were maintained since an overspecification is better than a underspecification.  

Results from four methods are shown in Table \ref{um_tab2}. The first one is the usual GEE method using the available data; the second is the weighted method (WGEE) using model (\ref{um_mod.perda}) for the weights; the third is the multiple imputation by chained equation (MIGEE) in the \emph{R} package \emph{mice}; and the fourth, labeled DRGEE, is the proposed doubly robust method using (\ref{um_mod.perda}) and (\ref{um_mod.cov}) for the weight and the covariate models, respectively. It was used an independent working correlation. 

\begin{table}
	\label{t:two}
	\begin{center}
		\scriptsize
		\caption{Regression Parameters for the Analgesia in Birth Data}\label{um_tab2}
		\begin{tabular}{rrrrrrrrrrrrrrrr}
			\hline
			\hline
			& \multicolumn{3}{c}{Available} &       & \multicolumn{3}{c}{WGEE} &       & \multicolumn{3}{c}{MIGEE} &       & \multicolumn{3}{c}{DRGEE} \\
			\cline{2-4}\cline{6-8}\cline{10-12}\cline{14-16}    Parameter & \multicolumn{1}{c}{Est} & \multicolumn{1}{c}{SE} & \multicolumn{1}{c}{P} & \multicolumn{1}{c}{} & \multicolumn{1}{c}{Est} & \multicolumn{1}{c}{SE} & \multicolumn{1}{c}{P} & \multicolumn{1}{c}{} & \multicolumn{1}{c}{Est} & \multicolumn{1}{c}{SE} & \multicolumn{1}{c}{P} & \multicolumn{1}{c}{} & \multicolumn{1}{c}{Est} & \multicolumn{1}{c}{SE} & \multicolumn{1}{c}{P} \\
			\hline
			INTERCEPT1 & 1.796 & 1.308 & 0.170 &       & 1.827 & 1.267 & 0.149 &       & 1.599 & 1.237 & 0.196 &       & 1.649 & 1.315 & 0.210 \\
			INTERCEPT2 & 3.302 & 1.314 & 0.012 &       & 3.261 & 1.310 & 0.013 &       & 3.105 & 1.242 & 0.012 &       & 3.069 & 1.351 & 0.023 \\
			TIME & -0.182 & 0.286 & 0.524 &       & -0.157 & 0.315 & 0.619 &       & -0.110 & 0.273 & 0.687 &       & -0.192 & 0.293 & 0.511 \\
			GROUP & -1.221 & 0.445 & 0.006 &       & -1.244 & 0.439 & 0.005 &       & -1.056 & 0.409 & 0.010 &       & -1.008 & 0.390 & 0.010 \\
			AGE & -0.066 & 0.035 & 0.057 &       & -0.072 & 0.033 & 0.027 &       & -0.062 & 0.031 & 0.046 &       & -0.071 & 0.035 & 0.045 \\
			DU & -0.362 & 0.158 & 0.022 &       & -0.354 & 0.168 & 0.035 &       & -0.372 & 0.156 & 0.017 &       & -0.340 & 0.166 & 0.040 \\
			OXYT(=2) & 0.727 & 0.554 & 0.189 &       & 0.712 & 0.554 & 0.199 &       & 0.954 & 0.542 & 0.078 &       & 0.821 & 0.531 & 0.122 \\
			OXYT(=3) & 1.285 & 0.510 & 0.012 &       & 1.431 & 0.503 & 0.004 &       & 1.410 & 0.488 & 0.004 &       & 1.468 & 0.475 & 0.002 \\
			\hline
		\end{tabular}
	\end{center}
\end{table}

TIME effect is non significant for all the four methods. All methods provide the same conclusion for effects of GROUP. The negative effect for GROUP means that the chance of women feel mild pain is lower among the group receiving the remifentanil compared to the peridural group (the estimated odds is $e^{-1.008}=0.365$ in the doubly robust method). All methods also agree with respect to the effect of DU. That is, for each increase of $1$ cm of uterine dilation the chance of the parturient feel mild pain decreases (in the DRGEE it is $e^{-0.340}=0.712$, for example). It can be noticed that $p$-value for AGE effect goes from a non-significant of $0.057$ in the standard GEE to a significant one in DRGEE, as well as for the other two missing data approaches. The conclusion is that older women have lower chance of experiencing mild pain than young women. For the OXIT covariate all methods reached the same conclusion.

\section{Discussion}\label{um_discuss}

When longitudinal ordinal data are of interest, doubly robust estimator is a nice alternative. Doubly robust method combines ideas from weighting and imputation and has been applied elsewhere for estimation of means, causal inference, and in the longitudinal setting for binary response data (\cite{bang2005}, \cite{carpenter2006}, \cite{seaman2009}, \cite{chen2011}, \cite{li2013}). However, as far as we know, it has not been investigated for the longitudinal ordinal case. A doubly robust estimator is attractive in the sense that it needs only the correct specification of at least one of the models, but not necessarily both. Simulation results have indicated that, when at least the covariate model or missing data model is correct, the doubly robust estimators are consistent and present small-sample bias comparable to single robust alternatives MIGEE or WGEE. The proposed method presented good coverage probabilities, as well as its competitors but with a slight larger variance than multiple imputation. Simulation results also indicated that the bias of doubly robust estimators when both the covariate model and the missing data model are incorrect was the same magnitude of misspecified WGEE or MIGEE. We hope that, in practical applications, none of the predictive models would be grossly misspecified and then the proposed estimator would have a great potential of reducing the bias if the MAR assumption is correct.

When the assumed independent working correlation structure differs from the true underlying structure, there is no price to pay in terms of the consistency and asymptotic normality of $\boldsymbol{\beta}$, but such a poor choice may result in loss of efficiency \citep{molenberghs2005}. However, modeling of the association structure in the presence of missing data remains a challenge, specially with longitudinal ordinal data, because there is no direct way of modeling the association parameters.
Future research involves the investigation of the impact of other association structures in the doubly robust estimates. 

In the proposed doubly robust estimator marginal means were modeled by cumulative logits. This implies a  proportional odds model that in some cases may not be valid. Another future possible extension of the proposed model is, therefore, to allow non-proportional odds for a subset of the explanatory variables \citep{peterson1990}.

\addcontentsline{toc}{section}{References}
\bibliographystyle{authordate1}
\bibliography{biblio}

\newpage
\appendix
\section{Appendix}
\subsection{Asymptotic Variance}
To state the asymptotic properties of $\boldsymbol{\hat{\beta}}$, let\\
$\boldsymbol{S}_{1i}(\boldsymbol{\beta},\boldsymbol{\psi},\boldsymbol{\gamma})$ be the individual's contribution to the estimating equations for $\boldsymbol{\beta}$,\\
$\boldsymbol{S}_{2i}(\boldsymbol{\psi})$ be the individual's contribution to the  estimating equations for  $\boldsymbol{\psi}$, and\\
$\boldsymbol{S}_{3i}(\boldsymbol{\gamma})$ be the individual's contribution to the estimating equations for  $\boldsymbol{\gamma}$.

Define 
$\boldsymbol{\Gamma}(\boldsymbol{\beta},\boldsymbol{\psi},\boldsymbol{\gamma})=
E\left\{ \partial \boldsymbol{S}_{1i}(\boldsymbol{\beta},\boldsymbol{\psi},\boldsymbol{\gamma})/ \partial \boldsymbol{\beta}^T   \right\}$,
$\boldsymbol{I}_{12}(\boldsymbol{\beta},\boldsymbol{\psi},\boldsymbol{\gamma})=E\left\{ \partial \boldsymbol{S}_{1i}(\boldsymbol{\beta},\boldsymbol{\psi},\boldsymbol{\gamma})/ \partial \boldsymbol{\psi}^T   \right\}$,
$\boldsymbol{I}_{13}(\boldsymbol{\beta},\boldsymbol{\psi},\boldsymbol{\gamma})=E\left\{ \partial \boldsymbol{S}_{1i}(\boldsymbol{\beta},\boldsymbol{\psi},\boldsymbol{\gamma})/ \partial \boldsymbol{\gamma}^T  \right\}$, $\boldsymbol{I}_{2}(\boldsymbol{\psi})=E\left\{ \partial \boldsymbol{S}_{2i}(\boldsymbol{\psi})/ \partial \boldsymbol{\psi}^T   \right\}$,
$\boldsymbol{I}_{3}(\boldsymbol{\gamma})=E\left\{ \partial \boldsymbol{S}_{3i}(\boldsymbol{\gamma})/ \partial \boldsymbol{\gamma}^T   \right\}$, and
$\boldsymbol{Q}_i(\boldsymbol{\beta},\boldsymbol{\psi},\boldsymbol{\gamma})=\boldsymbol{S}_{1i}(\boldsymbol{\beta},\boldsymbol{\psi},\boldsymbol{\gamma})-
\boldsymbol{I}_{12}(\boldsymbol{\beta},\boldsymbol{\psi},\boldsymbol{\gamma})\boldsymbol{I}_{2}^{-1}(\boldsymbol{\psi})\boldsymbol{S}_{2i}(\boldsymbol{\psi})-
\boldsymbol{I}_{13}(\boldsymbol{\beta},\boldsymbol{\psi},\boldsymbol{\gamma})\boldsymbol{I}_{3}^{-1}(\boldsymbol{\gamma})\boldsymbol{S}_{3i}(\boldsymbol{\gamma})$.

\begin{theo}
	If either the missing data model or the covariate model is correctly specified, then
	\begin{equation}\label{vardr}
	n^{1/2}(\boldsymbol{\hat{\beta}}-\boldsymbol{\beta_0})\longrightarrow N(\boldsymbol{0},\boldsymbol{\Gamma}^{-1}(\boldsymbol{\beta}_0,\boldsymbol{\psi}_0,\boldsymbol{\gamma}_0)\boldsymbol{\Sigma}
	\left\{\boldsymbol{\Gamma}^{-1}(\boldsymbol{\beta}_0,\boldsymbol{\psi}_0,\boldsymbol{\gamma}_0)\right\}^T),
	\end{equation}
	where $\boldsymbol{\beta}_0$ is the true value of $\boldsymbol{\beta}$, $\boldsymbol{\psi}_0$ and $\boldsymbol{\gamma}_0$ are the probability limits of $\boldsymbol{\hat{\psi}}$ and $\boldsymbol{\hat{\gamma}}$, and \\ $\boldsymbol{\Sigma}=E\left\{\boldsymbol{Q}_i(\boldsymbol{\beta}_0,\boldsymbol{\psi}_0,\boldsymbol{\gamma}_0)\boldsymbol{Q}_i^T(\boldsymbol{\beta}_0,\boldsymbol{\psi}_0,\boldsymbol{\gamma}_0)    \right\}$.
\end{theo}

Inferences for $\boldsymbol{\beta}$ follows by replacing the unknown quantities in (\ref{vardr}) by its consistent estimators. We make use of ``generalized information equality'' \citep{pierce1982} that \\
$E\left\{ \partial \boldsymbol{S}_{1i}(\boldsymbol{\beta},\boldsymbol{\psi},\boldsymbol{\gamma})/ \partial \boldsymbol{\psi}^T  \right\}=-
E\left\{ \boldsymbol{S}_{1i}(\boldsymbol{\beta},\boldsymbol{\psi},\boldsymbol{\gamma})\boldsymbol{S}_{2i}^T(\boldsymbol{\psi})\right\}$, and \\
$E\left\{ \partial \boldsymbol{S}_{1i}(\boldsymbol{\beta},\boldsymbol{\psi},\boldsymbol{\gamma})/ \partial \boldsymbol{\gamma}^T  \right\}=-
E\left\{ \boldsymbol{S}_{1i}(\boldsymbol{\beta},\boldsymbol{\psi},\boldsymbol{\gamma})\boldsymbol{S}_{3i}^T(\boldsymbol{\gamma})\right\}$. Similarly \citep{robins1995},\\
$E\left\{ \partial \boldsymbol{S}_{2i}(\boldsymbol{\psi})/ \partial \boldsymbol{\psi}^T  \right\}=-
Var\left\{\boldsymbol{S}_{2i}(\boldsymbol{\psi})  \right\}$, and
$E\left\{ \partial \boldsymbol{S}_{3i}(\boldsymbol{\gamma})/ \partial \boldsymbol{\gamma}^T  \right\}=-
Var\left\{\boldsymbol{S}_{3i}(\boldsymbol{\gamma})  \right\}$.

The matrix $\boldsymbol{\Gamma}$ is replaced by $\boldsymbol{\hat{\Gamma}}=n^{-1}\sum_{i=1}^n\left\{ \partial \boldsymbol{S}_{1i}(\boldsymbol{\hat{\theta}})/\partial \boldsymbol{\beta}^T  \right\}$, and 
$\boldsymbol{\Sigma}$ by $\boldsymbol{\hat{\Sigma}}=n^{-1} \sum_{i=1}^n \left\{\boldsymbol{\hat{Q}}_i\boldsymbol{\hat{Q}}_i^T  \right\}$,
$\boldsymbol{\hat{Q}}_i=\boldsymbol{S}_{1i}(\boldsymbol{\hat{\theta}})-\boldsymbol{\hat{I}}_{12}(\boldsymbol{\hat{\theta}})\boldsymbol{\hat{I}}_2^{-1}(\boldsymbol{\hat{\psi}})\boldsymbol{S}_{2i}(\boldsymbol{\hat{\psi}})-\boldsymbol{\hat{I}}_{13}(\boldsymbol{\hat{\theta}})\boldsymbol{\hat{I}}_3^{-1}(\boldsymbol{\hat{\gamma}})\boldsymbol{S}_{3i}(\boldsymbol{\hat{\gamma}})$,
$\boldsymbol{\hat{I}}_{12}(\boldsymbol{\hat{\theta}})=n^{-1}\sum_{i=1}^n\left\{ \partial \boldsymbol{S}_{1i}(\boldsymbol{\hat{\theta}})/ \partial \boldsymbol{\psi}^T   \right\}$,
$\boldsymbol{\hat{I}}_{13}(\boldsymbol{\hat{\theta}})=n^{-1}\sum_{i=1}^n\left\{ \partial \boldsymbol{S}_{1i}(\boldsymbol{\hat{\theta}})/ \partial \boldsymbol{\gamma}^T \right\}$,
$\boldsymbol{\hat{I}}_{2}(\boldsymbol{\hat{\psi}})=n^{-1}\sum_{i=1}^n\left\{ \partial \boldsymbol{S}_{2i}(\boldsymbol{\hat{\psi}})/ \partial \boldsymbol{\psi}^T \right\}$, \\
$\boldsymbol{\hat{I}}_{3}(\boldsymbol{\hat{\gamma}})=n^{-1}\sum_{i=1}^n\left\{ \partial \boldsymbol{S}_{3i}(\boldsymbol{\hat{\gamma}})/ \partial \boldsymbol{\gamma}^T \right\}$.

The proof is similar to \cite{chen2011} and is omitted here.

\end{document}